# Stochastic Amplification of Fluctuations in Cortical Up-States

Jorge Hidalgo[1], Luís F. Seoane[2,3], Jesús M. Cortés[4,5], Miguel A. Muñoz[1,*]

[1] *Departamento de Electromagnetismo y Física de la Materia e Instituto de Física Teórica y Computacional Carlos I. Universidad de Granada, E-18071 Granada, Spain.*

[2] *Bernstein Center for Computational Neuroscience, Technische Universität Berlin, Germany.*

[3] *Current address: ICREA-Complex Systems Lab, Universitat Pompeu Fabra, Dr Aiguader 88, 08003 Barcelona, Spain.*

[4] *DECSAI: Departamento de Ciencias de la Computacion e Inteligencia Artificial. Universidad de Granada, E-18071 Granada, Spain.*

[5] *Current address: IKERBASQUE: The Basque Foundation for Science. BIOCRUCES: Health Research Institute. Hospital Universitario de Cruces Plaza de Cruces, s/n 48903. Barakaldo, Bizkaia. Pais Vasco, Spain*

[*] E-mail: mamunoz@onsager.ugr.es

## Abstract

Cortical neurons are bistable; as a consequence their local field potentials can fluctuate between quiescent and active states, generating slow $0.5 - 2$ Hz oscillations which are widely known as transitions between Up and Down States. Despite a large number of studies on Up-Down transitions, deciphering its nature, mechanisms and function are still today challenging tasks. In this paper we focus on recent experimental evidence, showing that a class of spontaneous oscillations can emerge within the Up states. In particular, a non-trivial peak around 20 Hz appears in their associated power-spectra, what produces an enhancement of the activity power for higher frequencies (in the $30 - 90$ Hz band). Moreover, this rhythm within Ups seems to be an emergent or collective phenomenon given that individual neurons do not lock to it as they remain mostly unsynchronized. Remarkably, similar oscillations (and the concomitant peak in the spectrum) do not appear in the Down states. Here we shed light on these findings by using different computational models for the dynamics of cortical networks in presence of different levels of physiological complexity. Our conclusion, supported by both theory and simulations, is that the collective phenomenon of "stochastic amplification of fluctuations" – previously described in other contexts such as Ecology and Epidemiology– explains in an elegant and parsimonious manner, beyond model-dependent details, this extra-rhythm emerging only in the Up states but not in the Downs.

## Introduction

The cerebral cortex exhibits spontaneous activity even in the absence of external stimuli. Deciphering its oscillations and their correlates to behavior and function are major challenges in Neuroscience [1, 2]. Thus, for instance, high-frequency neural activity in the $\beta$ and $\gamma$ ranges $(10 - 100$ Hz) has been related to a plethora of cognitive tasks including action, perception, memory, or attention [1]. On the other hand, slow $\delta$ waves $(0.5 : 2$ Hz) are preponderant during the deepest stages of sleep, under anesthesia, or during quiet wakefulness [3–5], and may play an important role in neural plasticity and in the consolidation of new memories [6]. Finally, changes in the pattern of global activity are associated with brain-state transitions such as sleep-wake or to pathologies such as epilepsy [7]. Remarkably, very similar patterns of activity have been observed *in vitro* as well; both, coherent oscillations in the beta-gamma ranges and slow oscillations have been reported in brain slices [8–11], what suggests that these spontaneous oscillations are intrinsic to the dynamics of cortical networks.

These slow oscillations appear in the form of *Up-and-Down states* in which a large fraction of neurons alternate coherently between two different stable membrane-potential states: the quiescent *Down state* –with a high degree of hyper-polarization and very low activity– and the depolarized *Up state* –with high synaptic and spiking activity– [12]. The coherent (though non-periodic) -alternation between Up- and Down- states gives rise to Up-and-Down transitions, resulting in low-frequency $\delta$ waves [13]. The function and role of such transitions at the global network level are not fully understood (see [14] and references therein). The origin of such a bistability in the cortex dynamics has been argued to rely either on intrinsic neuronal features [9,15,16] or on network-level properties [17–19]. Even if its nature is not universally agreed upon, most of the existing computational models for cortical Up-and-Down states feature network rather than cellular mechanisms [13]. Here, we will focus on network models in which the cortex bistability emerges as a collective network phenomenon.

Existing computational models for network bistability involved some regulatory mechanism such as short time synaptic depression [18, 20, 21] or the presence of inhibitory populations of neurons [16,17,22]. Any of these ingredients (repressors) provides a negative feedback mechanism able to control the overall level of activity generated by self-excitation, allowing for the network to self-regulate. Generically, network models including activator/repressor dynamics may exhibit two different possible outputs, with low and high levels of activity, respectively. Although it is also possible to switch in the absence of noise between these two levels (eg. through a limit cycle), most of the previous models incorporate noise-induced Up-Down transitions, and in this paper we follow this strategy.

Given the apparent dichotomy between slow and high-frequency oscillations and their distinct cognitive correlates and function, the empirical finding that slow and fast rhythms may coexist might sound surprising but it has been shown to occur by different authors. Firstly, Steriade et al. found that high-frequency oscillations occurred within the active intervals of slow oscillations [23]. In similar experiments, Mukovski et al. [24], Fujisawa et al. [25], and more recently Compte and coauthors [26] have shown that high-frequency oscillations –in the 10-80 Hz range– develop within the Up intervals of Up-and-Down states. In particular, the power spectrum of such oscillations develops a pronounced peak at some frequency in the $\beta$-band –between 20 and 30 Hz– together with a substantial increase in the spectral power all along the $\beta/\gamma$ range. Remarkably, no similar peak has ever been observed in Down states [25, 26].

Another remark acknowledged by Compte et at. in [26] is that, while measurements of local field potentials in the Up state reveal robust oscillations in the $\beta/\gamma$, individual membrane potentials at the intracellular level do not show any trace of similar oscillations in that frequency band. This suggests, on the one hand, that high-frequency oscillations are a collective phenomenon emerging at the network level and, second, that there is no global synchronization (frequency locking) of individual neurons to the systemic rhythm. Thus, individual neural rhythms and the global emerging rhythm are independent.

At the modeling side, several authors have before addressed some of these issues and computed, in particular, the power-spectrum of network oscillations. For instance, Kang et al. [27] studied a mean field model in the presence of noise. They performed an analytical calculation of the power spectrum of a Wilson-Cowan-like model with excitatory and inhibitory neurons and showed the emergence of a resonant peak at gamma frequency. In a similar model, Wallace et al. [28] made the noise variance to scale with the network size and derived analytically the power-spectrum showing that it is possible to have coexistence of high-frequency oscillations for the population without having oscillations for individual neurons. On the other hand, for spiking neural networks, Spiridon and Gerstner [29] showed that the noise accounting for network-size effects affected the power-spectrum of the population activity. Similarly, and by using a Fokker-Planck formalism, Mattia and Del Giudice [30, 31], described the time evolution of the average network activity in presence of size-effects noise, and analytically derived its power spectrum and their resonant peaks.

Even if much has been written and is known about neural oscillations, our goal here is to shed some more light on the previously discussed questions by studying general aspects, beyond modeling details, as well as a simple and general theory accounting in general for the above described phenomenology and, in particular, for the asymmetry between Up state and Down state power spectra. For this, we study two different network models, one mean field and the other a network of spiking neurons, and discern whether high-frequency collective oscillations exist within the Up and/or within the Down state, respectively. Some of our results coincide with existing ones, as those reported in the previous paragraph, but, using a unified approach, here we conclude that a phenomenon termed *stochastic amplification of fluctuations* which can operate during Up –but not Down– states explains all the observations above in a robust, precise, and parsimonious way.

## Materials and Methods

Hereafter, we present two different network models reproducing the dynamics of Up-and-Down states, one based on a mean-field single population model (Model A) and one based on a network of spiking-neurons (Model B). Our strategy is to keep models as simple as possible to uncover the essence of Up-and-Down states. The theory of stochastic amplification of fluctuations, aimed at accounting for the non-trivial phenomenology above beyond modeling details, is presented also in this section.

### Model A: Minimal model for Up-and-Down states

The simplest possible models for Up and Down states have a deterministic dynamics and characterize neural network activity by a global ("mean-field") variable, the population averaged firing rate (which is a proxy for measurements of local field potential). Different models including synaptic depression and/or some other regulatory mechanism such as inhibition, have been employed in the past to describe Up and Down states. We focus here on the model proposed by Tsodyks *et al.* [32, 33]) including activity-dependent short-term synaptic plasticity as the key regulatory mechanism. In the S1 we present results for a similar model with inhibition. In this context, Up and Down states correspond to fixed points of the deterministic dynamics with, respectively, high and low firing-rates. The deterministic model is described by the mean membrane potential, $v$, and the variable $u$ accounting for the strength of synaptic depression. This second variable mimics the amount of available resources (varying between 0 and 1) in the presynaptic terminal to be released after presynaptic stimulation, thus, the larger $u$ the more synaptic input arriving to the postsynaptic cell [32, 33]. The mean voltage grows owing to both external and internal inputs and decreases owing to voltage leakage. On the other hand, synaptic resources are consumed in the process of transmitting information and generating internal activity (providing a self-regulatory mechanism) and spontaneously recover to a target maximum value, fixed here to $u = 1$:

$$\begin{aligned}\dot{v} &= -\frac{v - V_r}{\tau} + \frac{w_{in}\mu u f(v)}{\tau} \\ \dot{u} &= \frac{1 - u}{\tau_R} - \mu u f(v),\end{aligned} \quad (1)$$

where $\tau = RC$ ($R$ membrane resistance and $C$ capacitance) and $\tau_R$ are the characteristic times of voltage leakage and synaptic recovery, respectively, $w_{in}$ is the amplitude of internal inputs, $V_r$ is the resting potential, and $\mu$ is the release fraction indicating the efficiency of synapses. The firing rate function, $f$, is assumed to depend on $v$ as $f(v) = \alpha(v - T)$ if $v \geq T$, where $T$ is a threshold value, and $f(v) = 0$ otherwise (i.e. it is a "threshold-linear" gain function). External inputs could also be added to the model, but they are irrelevant for our purposes here. Spontaneous transitions between these two stable states can also be described within this framework by switching-on some stochasticity. Possible sources of noise are network size effects, sparse connectivity, unreliable synaptic connections, background net activity, synapses heterogeneity, or irregular external inputs. An instance of this stochastic approach is the work of Holcman and Tsodyks [18] (see also [34]) where a noise term was introduced into the above mentioned mean-field model with synaptic depression. Indeed, adding uncorrelated Gaussian white noises, $\eta_v(t)$ and $\eta_u(t)$, of amplitude $\sigma_v$ and $\sigma_u$ respectively, to equation 1, converts them into a set of stochastic/Langevin equations [18]. While the noiseless version of the model presents bistability its noisy counterpart exhibits Up-and-Down states.

## Model B: Spiking-neuron network model for Up-and-Down states

Millman and coauthors [21] proposed an integrate-and-fire (neuron-level) generalization of the model above, including some additional realistic factors. These refinements allow us to compare the emerging results with empirical ones not only qualitatively but also quantitatively. The model (Model B, from now on) consists in a population of $N$ leaky integrate-and-fire neurons, each one connected by excitatory synapses with (on average) another $K$ of them, forming a random (Erdos-Renyi) network. Each neuron is described by a dynamical equation for its membrane potential $V_i$ (with $i \in \{1, ..., N\}$) in which $V_i$ increases owing to (i) external (stochastic) Poisson-distributed inputs arriving at rate $f_e$ and (ii) internal inputs from connected spiking pre-synaptic neurons, and decreases owing to voltage leakage (see S2 for further details). When a neuron membrane potential $V_i$ reaches a threshold value $\theta$ the neuron fires: $V_i$ is reset to $V_r$ and its dynamics is switched-off during a refractory period $\tau_{rp}$. When a (pre-synaptic) neuron fires, it may open –with probability $p_r$– each of the $n_r$ release sites existing per synapse, inducing a current in the corresponding postsynaptic neuron. External (resp. internal) inputs, $I_e(t)$ (resp. $I_{in}(t)$) are modeled by exponentials of amplitude $w_e$ (resp. $w_{in}$) and time constant $\tau_s$. Similarly to Model A a variable $U_{ij} \in [0, 1]$ (for neuron $i$ and release site $j$) such that the release probability is modulated by $U_{ij}$, i.e. $p_r \to p_r U_{ij}$, allows to implement short-time synaptic depression. $U_{ij}$ is set to 0 immediately after a release and recovers exponentially to 1 at constant rate, $\tau_R$ (see S2).

**Stochastic amplification of fluctuations (SAF)**

Following [35] (see also [36] for an earlier reference) consider a set of deterministic equations, $\dot{v} = g_v(v, u)$ and $\dot{u} = g_u(v, u)$, complemented respectively with additive Gaussian white noises $\eta_v(t)$ and $\eta_u(t)$, giving rise to a set of two Langevin equations. To analyze fluctuations around a fixed point $(v^*, u^*)$ of the deterministic dynamics, a standard linear stability analysis can be performed. Defining $x = v - v^*$ and $y = u - u^*$, one can linearize the deterministic part of the dynamics

$$\begin{aligned} \dot{x} &= a_{vv}x + a_{vu}y + \eta_v(t) \\ \dot{y} &= a_{uv}x + a_{uu}y + \eta_u(t), \end{aligned} \quad (2)$$

where $a_{zz'} = \partial g_z(v,u)/\partial z'$ ($z$ and $z'$ standing for either $v$ or $u$) are the elements of the Jacobian matrix, $A$, evaluated at the fixed point. The associated eigenvalues $\lambda_\pm$ can be written as $\lambda_\pm = \Gamma/2 \pm \sqrt{\Gamma^2/4 - \Omega_0^2}$ with $\Omega_0^2 = \det(A) = a_{vv}a_{uu} - a_{vu}a_{uv}$ and $\Gamma = \text{Tr}(A) = a_{vv} + a_{uu}$.

A useful tool to identify oscillations in noisy time-series is the power spectrum $P_x(w) = \langle |\tilde{x}(w)|^2 \rangle$, where $\tilde{x}(w)$ is the Fourier transform of $x(t)$ (similarly $P_y(w)$ for $y(t)$), and $\langle . \rangle$ stands for independent runs average. Fourier transforming equation 2, solving for $\tilde{x}(w)$ and $\tilde{y}(w)$, and averaging its squared modulus, we find

$$P_z(\omega) = \frac{\alpha_z + \sigma_z^2 \omega^2}{\left[\Omega_0^2 - \omega^2\right]^2 + \Gamma^2 \omega^2} \tag{3}$$

where $z$ stands for $x$ or $y$, and $\alpha_x = a_{vu}^2 \sigma_y^2 + a_{uu}^2 \sigma_x^2$, $\alpha_y = a_{uv}^2 \sigma_x^2 + a_{vv}^2 \sigma_y^2$. For small noise amplitudes both of the power spectra exhibit maxima near

$$\omega_0 = \sqrt{\Omega_0^2 - \Gamma^2/2} = \sqrt{-a_{vu}a_{uv} - (a_{vv}^2 + a_{uu}^2)/2} \tag{4}$$

where the denominator has a minimum if $\omega_0$ is a real number. To have a real $\omega_0$ requires that both $a_{vu}$ and $a_{uv}$ are non-vanishing and of opposite sign; when this happens, both eigenvalues of $A$ are complex (see S3). As we shall see in what follows this condition is fulfilled for Up- but not for Down states. Finally, let us underline that $\omega_0$ does not depend on the noise amplitude.

The presence of a non-trivial peak in the spectrum of fluctuations reflects the existence of quasi-cycles of a leading characteristic frequency, coexisting with many other frequencies, and producing a complex oscillatory pattern. Notice that, even if the peak location $\omega_0$ is noise independent (as long as the noise amplitude does not vanish) the very presence of a peak is a noise induced effect: in the noiseless limit the system reaches a fixed point. The phenomenon we have just described –termed stochastic amplification of fluctuations (SAF)– has been recently put forward in the context of population oscillations in Ecology [35] (see also [36]) has also been claimed to be relevant in various other areas, such as Epidemiology [37]. SAF requires the presence of some noise source acting on top of the underlying deterministic stable fixed point with complex eigenvalues $\lambda_\pm$, i.e. the relaxation towards the stable fixed point should be in the form of damped oscillations (this is, it is a "focus") with a not too small damping frequency (details are explained in S3). Noise "kicks" the system away from the fixed point, and amplifies predominantly some frequency which –surprisingly enough– turns out to be *different* from the characteristic frequency of the deterministic damped oscillations (see S3). It is also noteworthy that a set of at least two coupled equations is required to have complex eigenvalues, and hence, too simplistic models in terms of only one effective variable, cannot give raise to SAF. Also, if the equations become decoupled (as it turns out to be the case for Down-states) the eigenvalues become real and the possibility of stochastic amplification is lost.

# Results

## Model A

Time-series produced by numerical simulations of such a Model A are shown in Fig. 1. Depending on the noise amplitude different outputs are produced. For low noises, either an Up state (with a high firing rate) or a stable Down state (with mean $v$ close to the resting potential, and therefore with a vanishing firing rate, and mean $u$ close to unity) coexist (converging into one or the other depends on the initial conditions). For larger noise Up-and-Down transitions are induced and Up-and-Down states emerge.

By performing a linear stability analysis equation 1 of as described above, we have measured the power-spectrum $P(w)$, both analytically and numerically, at either the Up state and the Down state. The deterministic Up-state fixed point turns out to be a focus, with complex eigenvalues, satisfying the conditions for the existence of a non-trivial peak in the power spectra for both $v$ and $u$. On the other hand, the Down-state fixed point (owing to the vanishing firing rate and, therefore, to the absence of crossed coupling terms ($a_{vu} = a_{uv} = 0$ in Eq.(2)) is a node with real eigenvalues and, consequently, there is no non-trivial peak in the power-spectrum.

These results are illustrated in Fig. 2. Observe (**i**) the perfect agreement between analytical and numerical results in all cases, (**ii**) the presence of a peak (around 1.6 Hz) for the $v$ power spectrum in the Up state (note that this rhythm is much faster than that of the Up-and-Down transitions, see Fig. 1), as well as (**iii**) the absence of similar peaks for the Down-state, and finally, (**iv**) the presence of a $w^{-2}$ tail in all power spectra. Very similar plots can be obtained –in analogy with measurements in [26]– in the Up-intervals within Up-and-Down states as well as for $u(t)$ as reported in S4.

Summing up, a mean-field single-population model in presence of short-term synaptic depression as the key regulatory ingredient reproduces Up-and-Down transitions, with a non-trivial peak in the up state power spectrum emerging as a consequence of the phenomenon of SAF. Numerical results are in full agreement with this theory, and consequently no analogous peak is found in Down states.

To test the generality of this hypothesis, we have also considered the mean-field dynamics of a simple model in presence of synaptic inhibition rather than synaptic depression (cf. S1). The model also exhibits Up-Down states transitions, with a non-trivial emerging peak in the Ups but not in the Downs, consistent with SAF. Remarkably, this supports that the phenomenon of SAF invoked here remains valid beyond the particular type of neuro-physiological mechanism for network self-regulation.

Despite this success, the strategy of resorting to simplistic mean-field models presents some undeniable drawbacks: (**i**) given the lack of a detailed correspondence with neuro-physiological realistic parameters it is not possible to *quantitatively* compare the results with experimental ones; (**ii**) noise is implemented in a poorly understood way; and (**iii**) last but not least, mean-field models do not allow for comparison of individual-neuron activity with collective rhythms, which is important to figure out whether single cells frequency-lock to emergent oscillations or not. Aimed at overcoming these difficulties, in the next section we present results for a network of spiking-neurons, Model B.

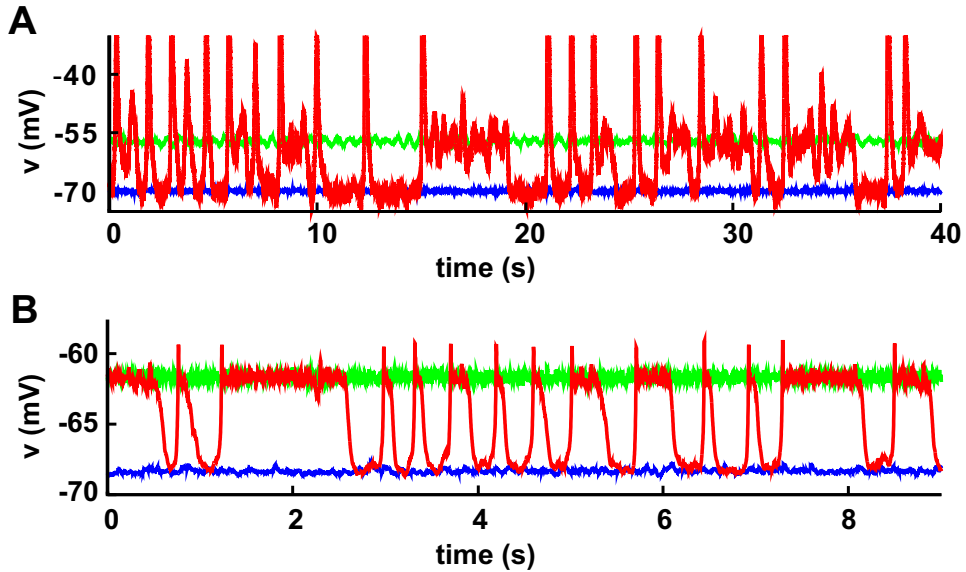

**Figure 1. Up and Down states and Up-and-Down transitions in two different network models. (A)** Model A (mean-field model) [18]: time-series for the membrane potential, $v(t)$. Observe the presence of two steady states lower one around $-70$ mV (Down-state/blue curve) and a larger one (Up state/green curve) at about $-55$ mV; these two are obtained for low noise amplitudes ($\sigma_v = 0.03$ mV/$\sqrt{\tau}$, $\sigma_u = 0.0004$ $1/\sqrt{\tau}$) and different initial conditions. Instead, the Up-and-Down state (red curve), corresponds to a high noise amplitude ($\sigma_v = 2.2$mV/$\sqrt{\tau}$, $\sigma_u = 0$). Note that, typically the Up-state intervals start with an abrupt spike which parallels empirical observations as discussed in [18]. Parameters have been fixed as in [18]: $\tau = RC = 0.05$ s, $\tau_R = 0.8$ s, $w_{in} = 12.6$ mV/Hz, $R = 0.5$, $T = -68.0$ mV, $V_r = -70$ mV, and $\alpha = 1.0$ Hz/mV. **(B)** Model B (network of spiking neurons) [21]: Time series of membrane potential. Curves and color code are as for Model A. For $p_r = 0.3$ the system exhibits Up-and-Down transitions, for larger (smaller) values as $p_r = 0.5$ ($p_r = 0.2$), it remains steadily in the Up (Down) state. Parameters have been fixed as in [21] : vesicles per synapsis $n_r = 6$, resting potential $V_r = -70$ mV, membrane threshold $\theta = -50$ mV, capacitance $C = 30$ pF, leakage characteristic time $\tau = RC = 0.02$ s, synaptic recovery time $\tau_R = 0.1$ s, signal time decay $\tau_s = 0.005$ s, refractory period $\tau_{rp} = 0.001$ s, input amplitudes $w_{in} = 50$ pA, $w_e = 95$ pA, and external driving rate $f_e = 5$ Hz.

## Model B

We have scrutinized Model B by numerically integrating the corresponding integrate-and-fire stochastic equations on sparse random networks as well as on regular networks. Parameters are fixed –mostly as in [21]– to neuro-biologically realistic values (see Fig. 1). We compute numerically membrane-potential and synaptic-resource time-series for each individual neuron as well as for the network as a whole. The release probability, $p_r$, is kept as a control parameter [32]: for intermediate values as $p_r = 0.3$ the system exhibits Up-Down transitions as illustrated in Fig. 1; for larger values (e.g. $p_r = 0.5$) it remains steadily in the Up state, while for sufficiently low ones ($p_r = 0.2$) only Down states are observed (see Fig. 1).

The power-spectrum $P(w)$ of the membrane potential time-series is illustrated in Fig. 2 (green for the Up state, blue for the Down one, both in linear and in double-logarithmic scale). Very similar plots can be obtained –in analogy with measurements in [26]– in the Up-intervals within Up-and-Down states as well as for $u(t)$ as reported in S4. In the Up state, the spectrum exhibits a sharp peak at a frequency around $\sim 20$ Hz, together with the expected power-law decay. On the other hand, the power spectrum for Down states lacks a similar peak. In analogy with the mean-field model in the previous section, there is a significant enhancement of the power-spectrum for Up vs Down states in the whole $\beta - \gamma$ range. However, on the contrary to the model above –giving the more detailed neuron-level modeling and the use of realistic parameter values– results can be *quantitatively* compared with empirical findings. Indeed, observe that, in remarkable accordance with the experimental observations in [26] (see, e.g. Fig. 1D in [26]) the peak in the Up state spectrum lies at frequencies in the $\beta2$-range, between 20 and 30 Hz. Let us remark that no parameter fine-tuning has been required to achieve this result.

Furthermore, Millman *et al.* showed in [21] that Up-and-Down states in Model B are robust against addition of fast AMPA currents, NMDA currents and (moderate) inhibition, more structured (small-world) network topologies, as well as voltage-dependent membrane resistance. Also, the non-trivial peak of the power-spectra and the associated spectral power enhancement in the $\beta/\gamma$ range for Up states, together with the absence of similar traits for Down states, are robust features against the extensions of the model we have scrutinized.

We have also analyzed time-series of individual neurons and compared their individual rhythms to that of the global, mean-field $v(t)$. Fig. 3 (left) shows that individual neurons do follow the global trend in Up-and-Down states: global high (resp. low) average membrane potentials correspond to high (resp. low) firing rates at the individual neuron level. On the other hand, and contrary to naive expectations, within Up states (as well as within Up periods of up-and-down states) where collective quasi-oscillations for the global mean-field emerge, individual neurons do not lock themselves to such a collective rhythm; as shown in Fig. 3 (right) individual neurons fire at a much faster pace than that of the global rhythm.

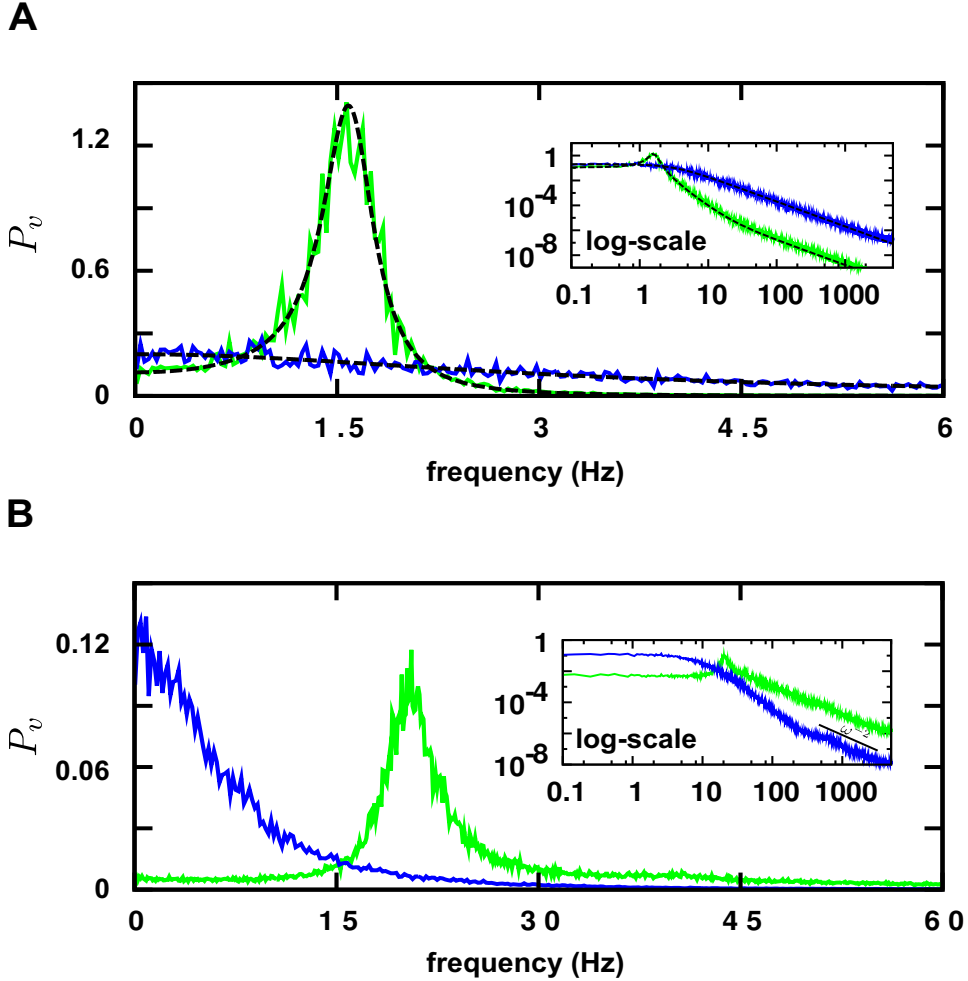

**Figure 2. Power spectrum of membrane potential $v(t)$ time-series in Up- and in Down states computed in Model A and Model B, respectively.** Histograms are normalized to unit area. The main plots show the power-spectra in linear scale: a pronounced peak appears for the Up state (green curve) around (A) $\approx 1.6$ Hz and (B) $\approx 20$ Hz. Instead, there is no track of similar peaks for Down states (blue curve). Observe the excellent agreement between simulation results (noisy curves) and analytical results for Model A, Eq.(3) (black dashed lines); for Model B a precise analytical prediction cannot be obtained. Insets represent analogous double logarithmic plots, illustrating in all cases the presence of $w^{-2}$ tails.

Actually, a histogram of the inter-spike intervals for all neurons in the network (shown in Fig. S5-1) has an averaged value $\approx 17$ ms, corresponding to a frequency $f \approx 60$ Hz. Therefore, given that the peak-frequency of the collective quasi-oscillations is located around 20Hz each neuron fires on average 3 times before a cycle of the collective rhythm is completed. The same result has been achieved by analyzing the power-spectrum for individual neurons, which turns out to exhibit a peak around $f \approx 60$ Hz and no sign of power enhancement in the $20 - 30$Hz band (see Fig. S5-2).

To firmly establish the correspondence between the just-described phenomenology for Model B and SAF we need to write down a set of effective Langevin equations, analogous to Eq.(1) for the global, network-averaged, variables and compute power-spectra from them. For a network of finite size, this can not be done in an exact way. However, as detailed in S2, the Fokker-Planck equation for the probability distribution of any *individual-neuron* membrane potential $V_i$ in Model B can be easily written down for infinite networks [21]. The network-averaged firing rate, $f$, appears explicitly in such an equation, and needs to be self-consistently determined: $f$ has to coincide with the outgoing probability flux, i.e. the fraction of neurons overcoming the threshold $\theta$ per unit time in the steady state [21]. By scrutinizing such a Fokker-Plank equation it is straightforward to see that individual neurons, follow an oscillatory pattern in which each of them is progressively charged and then fires at a pace that coincides with the (numerically determined above) rhythm of individual neurons. No track of SAF can be seen at this individual-neuron level.

In order to have an equation for the collective rhythms, we have taken the previous Fokker-Planck equation and from it computed the network-averaged membrane potential (needed to scrutinize the possible existence of SAF) at a network level, defined as

$$v(t) \equiv \int_{V_r}^{\theta} V P(V,t) dV. \tag{5}$$

and similarly, the network-averaged synaptic depression variable $u(t)$. As shown in the S6 they

obey

$$\begin{aligned}\dot{v} &= -(\theta - V_r)f(t) - \frac{v - V_r}{RC} + V_e f_e + KuV_{in}f(t) + DP(V_r, t) \\ \dot{u} &= \frac{1-u}{\tau_R} - p_r u f(t).\end{aligned} \qquad (6)$$

In the first equation $(\theta - V_r)f$ describes the average potential reduction owing to resetting, $-\frac{v-V_r}{RC}$ is the average leakage, $V_e f_e$ and $KuV_{in}f$ (with values of constants detailed in S2 and caption of Fig. 1) stand for the average external and internal charging, respectively, and $DP(V_r)$ is proportional to the fraction of neurons in the resting state. The two terms in the second equation describe average recovering and consumption of synaptic resources respectively.

Eq.(6), valid for infinitely large networks, are deterministic equations. Instead, for any finite network of size $N$, with finite connectivity and finite number of release sites, the former is no longer true: $f$ becomes a stochastic variable fluctuating around its averaged value. Something similar happens with the fraction of neurons at resting value, $P(V_r, t)$ appearing in Eq.(5).

Consequently, writing $f(t)$ and $P(V_r, t)$ as deterministic functions (depending on both variables, $v$ and $u$) plus a noise (fluctuating part), Eq.(6) becomes a set of Langevin equations, from which power spectra could be computed. However, determining analytically the functional dependence of $f(t)$ and $P(V, r, t)$ on $v$ and $u$ for finite values of $N$ (which is necessary to perform the stability analysis) is not feasible. Owing to this, we have resorted to a numerical evaluation of such dependences. Simulation results show that $P(V_r, t)$ hardly departs from its infinite $N$ limit value, and hence its variability can be neglected for all purposes here. Instead, $f$ depends strongly on $v$ and is almost independent of $u$; $f(v)$ can be approximated by a "threshold-linear gain function" –zero for $v < \theta'$ and linear when $v > \theta'$– as commonly used in the literature to approximate firing rates e.g. [18], plus a noise term, for both the Up and the Down state (see Fig. S6-1). It can also be verified that the amplitude of such a noise decreases with the square-root of the system size, as expected on the basis of the central limit theorem (see Fig. S6-2).

From 6, plugging in the approximate expression for $f(v)$ we can calculate analytically the fixed points of the deterministic dynamics, $v^*$ and $u^*$. Results agree reasonably well with numerically measured averaged values both in the Up and in the Down state. Having evaluated the deterministic fixed points we can follow a standard linear stability analysis as above, compute the stability matrix, the corresponding eigenvalues, and finally the power-spectra in the Up and in the Down state as detailed above (see S6 as well as S7). For the Up state the corresponding eigenvalues turn out to be complex (i.e. as explained above, $a_{uv}$ and $a_{vu}$ are both non-zero and of opposite signs, implying that $\omega_0$ is real) entailing a non-trivial peak in the power-spectrum located at $f_0 = 12.64$ Hz. This analytical prediction slightly deviates from the numerical results as reported in Fig. 2, exhibiting a peak at $f \simeq 20$ Hz. This deviation stems from the approximate nature of the present calculation. Developing a more precise analytical way to deal with finite networks remains an open and challenging task. On the other hand, for the Down state, the equations for $v$ and $u$ are essentially decoupled, eigenvalues are consequently real and, as a result, there is no peak in the power spectrum nor any significant enhancement of fluctuations.

In conclusion, we have shown that also for this more complex network model, an analytical (even if approximate) approach permits us to elucidate that the phenomenon of stochastic amplification of fluctuations is responsible for the non-trivial enhancement of fluctuations in the whole $\beta/\gamma$ range as well as the emergence of a peak in power spectra of Up states for a frequency in the $\beta 2$ band, around 20 Hz. Similar results do not hold for Down states.

## Discussion

Diverse computational models –with different levels of complexity– for Up-and-Down states have been introduced in the literature. Aimed at focusing on essential aspects of the Up-Down transitions, we choose here to scrutinize models as simple as possible. In particular we have studied two different models. The first one, Model A, is a "mean-field" model defined in terms of two global variables, equipped with some additional source of stochasticity. The second, Model B, is a neuron-level based network model. Both of them are described in terms of stochastic equations for membrane potentials as well as for a second variable modeling the dynamics of synaptic depression. A mechanism of activity-dependent (short-term) synaptic depression allows the system to generate negative feedback loops, ensuing self-regulation. Under these conditions, Up and Down states and Up-and-Down transitions emerge.

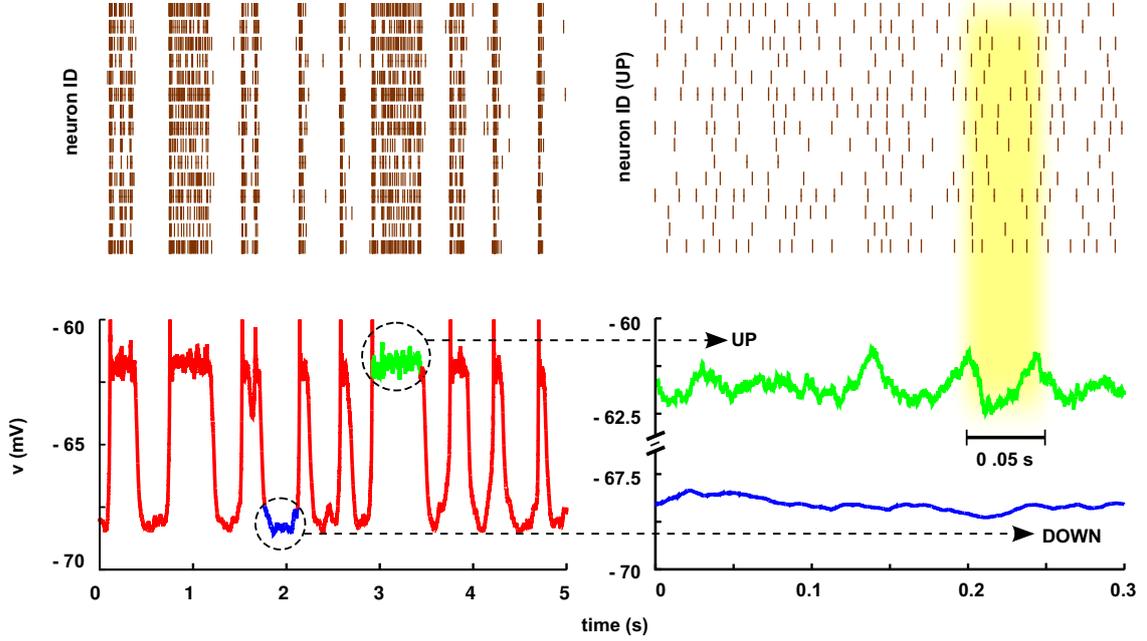

**Figure 3. Raster plots and average membrane potential in the spiking-neuron network model (Model B).** Left: (Top) Raster plot of 15 randomly chosen neurons (out of a total of $N = 1000$ neurons in the simulation). Sticks are plotted whenever a neuron spikes. (Bottom) Time-series of the network-averaged membrane potential in the same simulation. Comparison of the two left panels (both of them sharing the same time axis) reveals that individual neurons fire often during Up states, while they are essentially quiescent in Down-state intervals. Right: (Bottom) zoom of an Up interval (green curve) and of a Down interval (blue curve); while the Up state exhibits quasi-oscillations, the Down-state does not. (Top) Raster plot of 15 randomly chosen neurons during the Up state. Remarkably, their spiking frequency is not locked to the collective rhythm: it is about three times faster.

We first analyzed the simpler mean-field-like Model A describing activity at a global/macroscopic level, and then went on by introducing the spiking-neuron network Model B. For these, we have first performed computer simulations, confirming the existence of Up-and-Down states. To analyze fluctuations around either the Up or the Down state, power-spectra for the global (averaged) membrane potential –which is a proxy for experimentally measured local field potentials– have been computationally measured. They show similar phenomenology in all cases: in the Up state there is a non trivial peak at some frequency together with an overall enhancement of fluctuations in the whole $\beta/\gamma$ region, while no similar peak existing for Down states. These results are in excellent accordance with the experimental findings of diverse experimental groups –detailed in the Introduction– showing a similar enhancement of fluctuations under different experimental conditions in cortical Up states but never in Down states. Therefore, we conclude that existing models for Up-Down transitions succeed at reproducing realistic fluctuations in Up and Down states, as described in the Introduction.

The main contribution of the present work is to put forward that the empirically measured enhancement of fluctuations in Up states (as well as the lack of a similar effect in Down states) can be perfectly explained by the mechanism of "stochastic amplification of fluctuations". This mechanism consists in the resonant amplification of some frequencies in the spectra of stochastic systems when the corresponding fixed-point of its deterministic dynamics is a focus (i.e. in the infinite size limit the steady state fixed point has complex associated eigenvalues). The presence of any source of noise kicks the system away from the deterministic fixed point leading to a non-trivial power-spectrum. It is important to remark that (i) empirical measurements of local field potentials correspond to mesoscopic cortex regions, intrinsically affected by noise effects and hence, a stochastic description of them is fully justified, and that (ii) curiously enough, as explained here, the selected/amplified dominant frequency is *not* that of the deterministic damped oscillations towards the focus, as it could have been naively expected.

To firmly establish the correspondence between the non-trivial features of fluctuations observed empirically as well as in computer models for Up and Down states and the phenomenon of stochastic amplification, one needs to write down a deterministic equation for the network-averaged variables and complement it with a noise term, i.e. a Langevin equation. Writing down a Langevin equation for the global dynamics of Model A, which is already a mean-field model equipped with a noise term, is a trivial task. However, this is difficult for Model B, for which we have needed to resort to a more refined approach. In both cases, we have been able to construct analytical equations (exact) for Model A and (approximate) for Model B, study the associated power-spectra, and analytically confirm the presence of non-trivial peaks appearing owing to a stochastic amplification of fluctuations for Up states (which can be described by a fixed point with complex eigenvalues at a deterministic level) but not for Down states (with real valued deterministic eigenvalues).

While for the first-studied mean-field-like Model A the agreement between experimental results and theoretical predictions is only qualitative, for the more refined spiking-neuron network Model B, the accordance becomes also quantitatively good. Indeed, observe that, in remarkable accordance with the experimental observations in [26] (see, e.g. Fig. 1D in [26]) the peak in the Up state spectrum lies at frequencies in the $\beta 2$-range, between 20 and 30 Hz.

In any case, the reported phenomenon of stochastic amplification of fluctuations explains the emergence of quasi-oscillatory –with a typical dominant frequency and a broad power-spectrum– rhythms in the global-network activity within Up states as well as (owing to the absence of a significant firing rate) the absence of a similar effect for Down states. This explanation is robust beyond modeling specificities as confirmed by the finding that many model details can be changed without affecting the results and also by the fact that a very different model, based on inhibition rather than on synaptic depression, leads to identical conclusions. Using the jargon of excitable systems, we conjecture that any activator/repressor model –the repressor being, depression, inhibition or any other form of adaptation, is in principle able to induce SAF in Up states (but not in Down states) and consequently explain the non-trivial shape of power-spectra for cortical fluctuations.

Furthermore, we have shown that the mechanism of stochastic amplification of fluctuations operates for global variables but not for individual neurons. In the framework on the neuron-level based Model B, it is possible to compare the oscillatory behavior of single neurons with the network collective rhythms. We have explicitly shown that single neurons do not lock to the global collective rhythm emerging within Up states. Actually, single neurons fire at a much faster

pace –typically 3 times larger– than the collective oscillation period. This phenomenology, which perfectly accounts for empirical findings in [26] as reported in the Introduction, is similar to what has been called asynchronous-states or sparse-synchronization in which a collective rhythm – to which individual neurons do *not* lock– emerges (see [38, 39] for related, though different, phenomena). Observe that in the, so-called, "fast-oscillations", as described for instance in [38], the emerging global rhythm is much faster than individual neurons, while here it is the other way around.

In summary, Up and Down states as well as Up-and-Down transitions can be well described as collective phenomena emerging at a network level. They exhibit generically a set of highly non-trivial features which can be well captured by simple models, and perfectly accounted for by the mechanism of stochastic amplification of fluctuations.

## Acknowledgments

We acknowledge S. Johnson, J.A. Bonachela, and S. de Franciscis for useful discussions and a critical reading of earlier versions of the manuscript. We acknowledge financial support from the Spanish MICINN-FEDER under project FIS2009-08451, from Junta de Andalucía Proyecto de Excelencia P09FQM-4682. L.S. acknowledges the financial support of Fundación P. Barrié de la Maza and funding grant 01GQ1001A.

Note added: after completion of this work we became aware by the paper by Wallace et al. [28] now cited in the Introduction, in which similar effects to those described –including a explanation of the mechanism of SAF– where put forward, even if for a different (simpler) model. Interestingly enough, the collective oscillations found by Wallace et al. are much faster than individual neuron oscillations, which is just the opposite of what we (and the experiments in [26]) find for up states; obviously, these are two different instances of the same phenomenon: SAF.

# Supporting information for "Stochastic Amplification of Fluctuations in Cortical Up-states"

Jorge Hidalgo, Luís F. Seoane, Jesús M. Cortés, Miguel A. Muñoz

*Departamento de Electromagnetismo y Física de la Materia e Instituto de Física Teórica y Computacional Carlos I. Universidad de Granada, E-18071 Granada, Spain.*

## S1 Stochastic amplification in a excitation-inhibition mean-field model

Here we show that Stochastic Amplification of fluctuations can be found also in models for Up and Down states relying on populations of both, excitatory and inhibitory neurons. We consider the Wilson-Cowan-like model as described, for example, in [s1] (see also [s2] and [s3–s5] where a similar model has been recently studied). This is a mean-field like model, analogous in this sense to Model A, but with inhibition rather than depression as leading regulatory mechanism.

The model consists of two equations for the mean excitatory and inhibitory firing rates in the network:

$$\tau_e \dot{E} = -E + g(J_{ee}E - J_{ei}I + E_0) \tag{s1}$$

$$\tau_i \dot{I} = -I + g(J_{ie}E - J_{ii}I + I_0) \tag{s2}$$

with a threshold linear response function

$$g(x) = \begin{cases} 0 & x < T \\ \beta(x - T) & x \geq T \end{cases} \tag{s3}$$

where $T$ is a threshold parameter. For a wide range of parameter values these equations exhibit bistability: there is a stable fixed point with with low-activity regime (Down-state) and a second one with a non-vanishing firing rate and significant activity (Up-state). Adding Gaussian white noises to both equations (s1) and (s2), the system fluctuates and eventually may jump between the two fixed points. We have verified by means of computer simulations that indeed this model exhibits Up-and-Down states, that a non-trivial peak appears for fluctuations within Up states and not for Down states. The chosen parameters are shown in Table S1.

| Parameter | Value |
|---|---|
| $\tau_e, \tau_i$ | 0.01 s |
| $J_{ee}, J_{ie}, J_{ii}$ | 5 mV/Hz |
| $J_{ei}$ | 9 mV/Hz |
| $\beta$ | 0.5 Hz/mV |
| $T$ | 15 mV |
| $E_0$ | 10 mV |
| $I_0$ | 0 mV |

**Table S1.** Parameter values for the excitation-inhibition model presented in [s1].

Trajectories of the deterministic dynamics reveal spiral trajectories (i.e. damped oscillations) near the Up-state fixed point but not in the Down state (straight trajectories corresponding to real eigenvalues). Therefore, one can expect a non-trivial peak to appear in the Up-state power-spectrum but not in the Down one (see S3). This can be explicitly seen from a linear stability analysis, and indeed

$$A_{\text{down}} = \begin{pmatrix} -\frac{1}{\tau_e} & 0 \\ 0 & -\frac{1}{\tau_i} \end{pmatrix} \tag{s4}$$

$$A_{\text{up}} = \begin{pmatrix} -\frac{1}{\tau_e} + \beta J_{ee} & -\beta J_{ei} \\ \beta J_{ie} & -\frac{1}{\tau_i} - \beta J_{ii} \end{pmatrix}. \tag{s5}$$

$A_{\text{down}}$ is already diagonal in the Down state, with both real eigenvalues, therefore from equation (s14), $\omega_{0,\text{down}} \notin \mathbb{R}$ and hence, characteristic frequencies are not found in the power spectrum of fluctuations. On the other hand, eigenvalues are complex for the Up state, giving a peak in the $\beta$-range whose value is near $\omega_{0,\text{up}} = 200$ rad/s $= 31.8$ Hz. These analytical predictions are in excellent agreement with results of computer simulations for this model.

## S2 Model B of Millman et al. and its self-consistent solution

The model studied by Millman *et al.*, Model B, is defined by the following set of equations for the membrane potential $V_i$ of neuron $i$ and the synaptic utility $U_{ij}$ for each release site $j$:

$$\begin{aligned}
\dot{V}_i &= -\frac{V_i - V_r}{RC} + \frac{1}{C}\sum_k I_{e_i}^k(t) + \\
&\quad \frac{1}{C}\sum_{\substack{i'j' \\ \text{linking } i}}\sum_k \Theta(p_r U_{i'j'}(t_{s_{i'}}^k) - \zeta_{i'j'}^k) I_{\text{in}_{i'}}^k(t), \\
\dot{U}_{ij} &= \frac{1 - U_{ij}}{\tau_R} - \sum_k U_{ij}\Theta(p_r - \zeta_{ij}^k)\delta(t - t_{s_i}^k).
\end{aligned} \tag{s6}$$

where $\zeta_{ij}^k$ is a uniform random number in $[0,1]$ and $\Theta(x)$ the Heaviside step function. The first term in the r.h.s. of the first equation describes the leakage, the second is the sum over external inputs (Poisson distributed at rate $f_e$), and the third represents the internal currents arriving from (pre-synaptic) neuron $i'$ to (post-synaptic) neuron $i$ at every release site $j'$; there are $n_r$ release sites per synapse; $k$ runs over spikes, occurring at times $t_{s_i}^k$ for each neuron $i$.

The Fokker-Planck equation proposed in [s6] to describe this model in the limit of infinitely large system-size is

$$\begin{aligned}
\frac{\partial P(V,t)}{\partial t} &= -\frac{\partial F(V,t)}{\partial V} = \frac{\partial[\nu_d(V,t)P(V,t)]}{\partial V} + D\frac{\partial^2 P(V,t)}{\partial V^2} \\
&= \frac{\partial\left[(-\frac{V-V_r}{RC} + V_e f_e + KuV_{\text{in}}f)P(V,t)\right]}{\partial V} + \frac{1}{2}\left(V_e^2 f_e + Ku^2 V_{\text{in}}^2 f\right)\frac{\partial^2 P(V,t)}{\partial^2 V}.
\end{aligned} \tag{s7}$$

The drift (or deterministic) term in equation (s7) includes potential leakage and external plus internal input integration. The diffusion term stems from the Poisson-like nature assumed for both external and internal spikes ($K$ synapses per neuron; i.e. finite connectivity). $f$ stands for averaged firing rate and $V_{\text{in}} = p_r n_r w_{\text{in}} \tau_s / C$ and $V_{\text{e}} = w_{\text{e}} \tau_s / C$ are the mean increase in membrane potential from a single internal and external (exponential) input. Indeed, observe that the factor $\tau_s$ in the expressions $V_{\text{in/e}}$ comes from $\int_{t_s}^{\infty} e^{-(t-t_s)/\tau_s} dt = \tau_s$. In order to enhance the accuracy of the quantitative agreement between theoretical predictions and numerical results, we have improved this estimation of the global membrane potential increase per spike by taking into account that neurons are eventually reset and during their refractory period they do not integrate stimuli and the arriving inputs are interrupted. In this way, (see S7) the average input per spike becomes

$$\overline{V}_{\text{in/e}} = V_{\text{in/e}} \left[ 1 - f \tau_s \left( 1 - e^{-\frac{1}{f \tau_s}} \right) \right], \tag{s8}$$

which represents a significant change with respect to $V_{\text{in/e}}$ above.

Some remarks are in order:

- In the fully connected case $K = N$, assuming that internal input amplitudes are rescaled by the average connectivity (i.e. $w_{\text{in}} \to w_{\text{in}}/K$) in order to keep the total signal per spike constant, the internal noise disappears in the infinite size limit. In other words, the internal contribution to the diffusion term, proportional to $V_{\text{in}}$ stems from the finite connectivity of each individual neuron in sparse networks. Similarly, in the absence of external stochasticity, the external contribution to the diffusion term, proportional to $V_{\text{e}}$ would disappear for a homogeneously distributed excitation. If the two previous conditions hold, the dynamics becomes purely deterministic.

- Observe that only derivatives with respect to $V$, and not $u$, appear in equation (s7); this is because, the synaptic depression variable has been averaged also over release-sites, hence, in the limit of large $n_r$ it is replaced by its mean-field value which obeys

$$\dot{u}(t) = \frac{1-u}{\tau_R} - p_r f u. \tag{s9}$$

This replacement is accurate for large values of $n_r$ while otherwise it is just an approxi-mation.

Equation (s7) needs to be complemented with the boundary conditions $F(V_r, t+\tau_{rp}) = F(\theta, t)$ where $F(V,t) = \left(V_e^2 f_e + K u^2 V_{in}^2 f\right) \frac{\partial P(V,t)}{\partial V}$ is the flux at a given value of $V$, and $P(\theta, t) = 0$, representing the fact that neurons at threshold are instantly reset to the resting-potential $V_r$, and kept inactive for a refractory period $\tau_{rp}$ [s6, s7]. The firing rate, $f$, is computed as the outgoing probability flux, i.e. the fraction of neurons overcoming $\theta$ per unit time, $f(t) = F(\theta, t)$.

As the dynamics depends on the probability flux $f$, which on its turn is fixed by the overall dynamics, the Fokker-Planck equation needs to be solved self-consistently. This can be done numerically (Euler-implicit method) giving results in agreement with those in [s6]: there are two different stable states for the probability distribution (see figure S2).

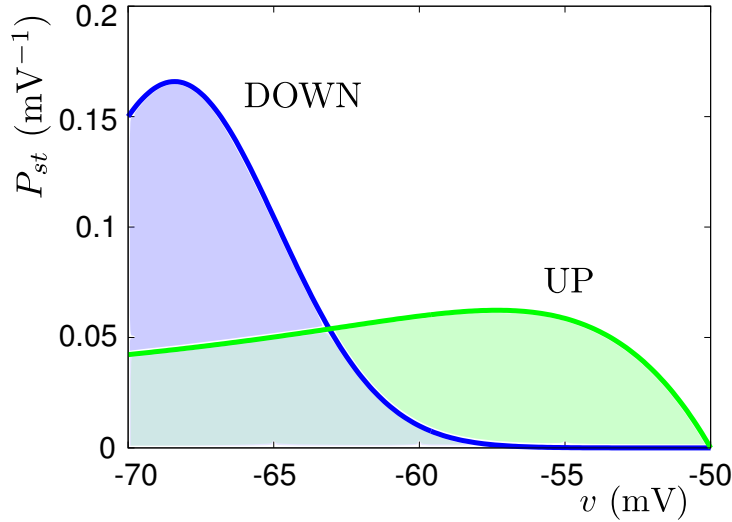

**Figure S2.** Solutions for the membrane potential distributions described by equation (s7). In the Down-state, membrane potentials are closer to $V_r$, and the slope in $\theta$ gives a low firing rate $f = 0.00022$ Hz, while in the Up-state, potentials raise up, giving $f = 74.9$ Hz.

## S3 Conditions for Stochastic amplification

Let us consider the stability (Jacobian) matrix, $A$, of a two-variable system and let $\lambda_\pm$ be its associated eigenvalues. In general, they can be written as complex numbers

$$\lambda_\pm = \lambda_\pm^R + i\lambda_\pm^I. \tag{s10}$$

As $A$ is a real matrix, its determinant and its trace are both real. This imposes some constraints on the eigenvalues: $\text{Tr}(A) = \lambda_+^R + \lambda_-^R + i(\lambda_+^I + \lambda_-^I) \in \mathbb{R}$ and hence

$$\lambda_+^I = -\lambda_-^I \equiv \lambda^I. \tag{s11}$$

Similarly, $\det(A) = \lambda_+^R \lambda_-^R - \lambda_+^I \lambda_-^I + i(\lambda_+^R \lambda_-^I + \lambda_-^R \lambda_+^I) \in \mathbb{R}$, and therefore

$$\lambda_+^R = \lambda_-^R \equiv \lambda^R \quad \text{if } \lambda^I \neq 0. \tag{s12}$$

As shown in the main text the power-spectrum can be expressed as

$$P(\omega) = \frac{\alpha_z + \sigma_z^2 \omega^2}{[\text{Det}(A) - \omega^2]^2 + (\text{Tr}A)^2 \omega^2}, \tag{s13}$$

which has a maximum around

$$\omega_0 = \sqrt{\det(A) - (\text{Tr}A)^2/2} = \sqrt{-\frac{1}{2}(\lambda_+^2 + \lambda_-^2)} \tag{s14}$$

where the denominator vanishes, provided that $\omega_0$ is real. Taking into account equations (s11) and (s12),

$$\omega_0 = \sqrt{(\lambda^I)^2 - (\lambda^R)^2} \tag{s15}$$

which provides a direct way to compute $\omega_0$. In particular, observe that $|\lambda^I| > |\lambda^R|$ is a necessary and sufficient condition for a non-trivial maximum to exist, and hence, the system does not exhibit Stochastic amplification if $\lambda^I$ is zero or not sufficiently large. Notice that, if $A$ is diagonal (i.e. the two equations become decoupled), $\lambda^I = 0$ and no stochastic amplification can occur. Indeed, it suffices that only one of the non-diagonal terms of $A$ is zero to rule out stochastic amplification.

Stochastic amplification of fluctuations occurs when the deterministic system falls with damped oscillations (spiral decay towards the focus, as corresponds to complex eigenvalues). Noise perturbs trajectories, kicking them away from the focus and sustaining oscillations. It is noteworthy that the selected oscillation frequency does *not* coincide with that of the transitory deterministic dynamics, $\omega^* = |\lambda^I| = \sqrt{\det(A) - \text{Tr}(A)^2/4}$, i.e. $\omega_0 \neq \omega^*$.

## S4 Power spectrum of fluctuations for the synaptic depression variable

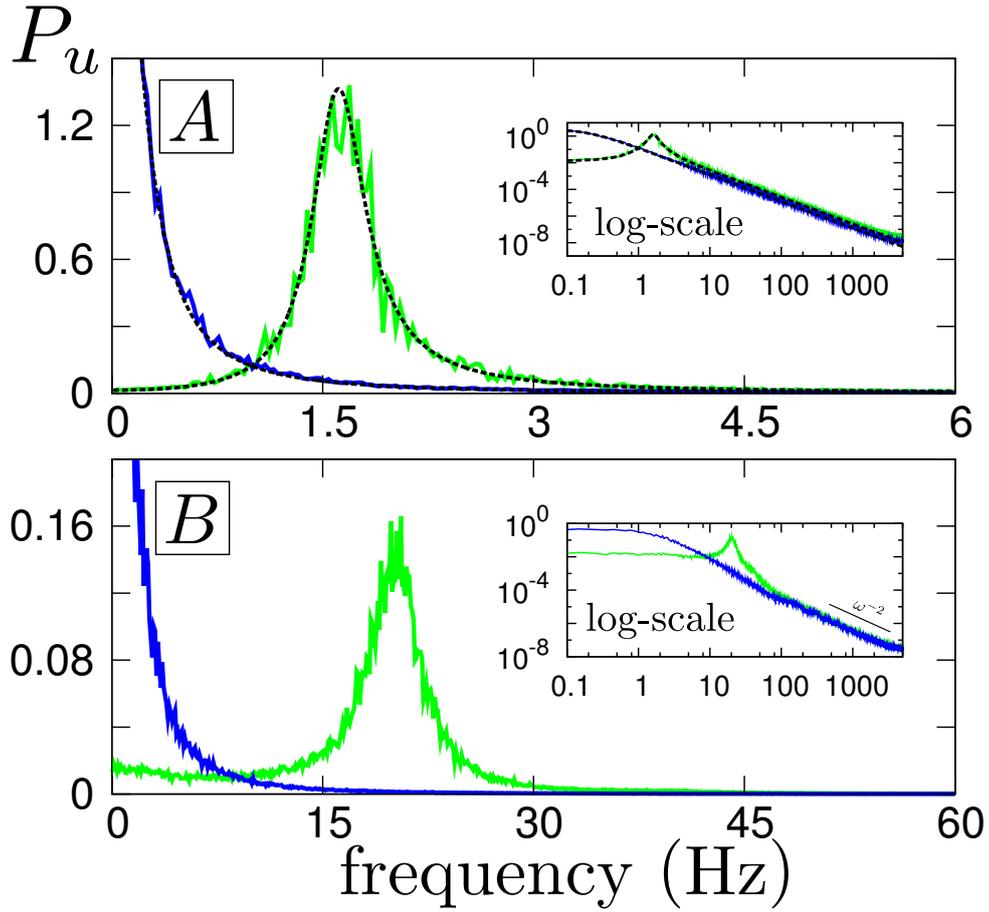

**Figure S4.** Power spectrum for the variable synaptic depression variable, $u$, in Up and in Down states for (A) Model A and (B) Model B, respectively. Plots are normalized to unit area. As in Fig. 2 of main text, similar peaks appear for Up (green curves) but not for Down (blue curves): (A) $\approx 1.6$ Hz and (B) $\approx 20$ Hz. Insets present power spectra in double logarithmic scale; spectra exhibit a $\omega^{-2}$ tail indicating the presence of many different scales.

# S5 Characteristic frequencies for individual neuron membrane potentials

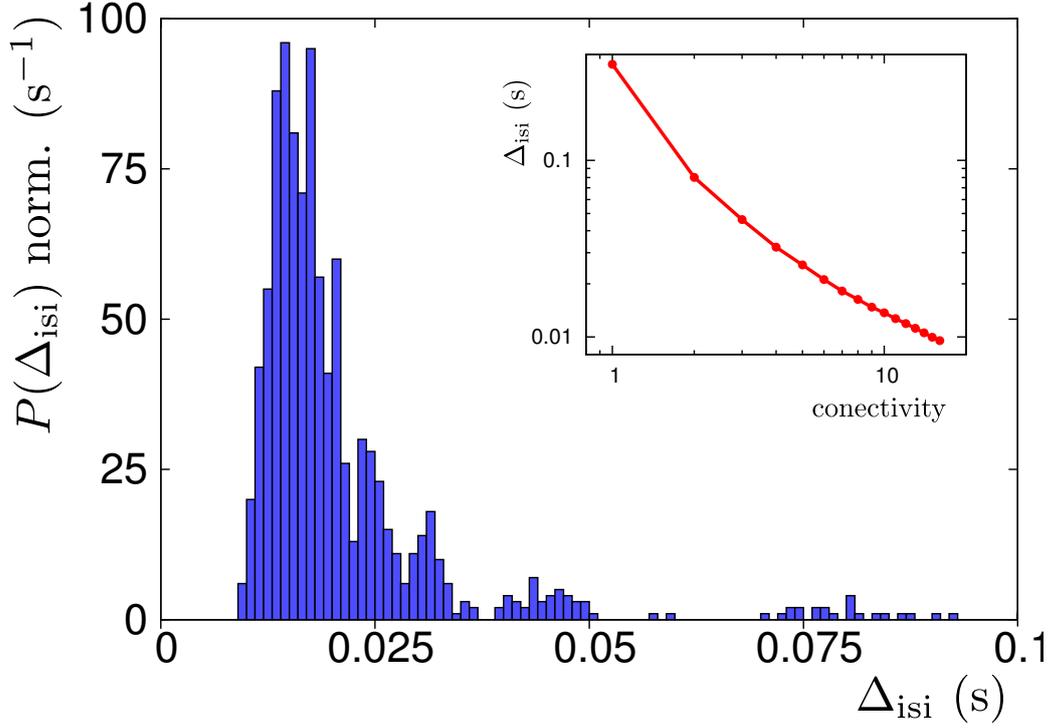

**Figure S5-1.** Probability distribution of the inter-spike-intervals in the random network; its average gives $\langle \Delta_{\text{isi}} \rangle \approx 17$ ms, corresponding to a mean firing rate of $f \approx 60$ Hz. This result perfectly agrees with mean firing rate of Fig. S5-2. Heterogeneity in the average inter-spike-intervals stems from the different connectivity degrees, as illustrated in the inset. In the latter, we show the average value of $\Delta_{\text{isi}}$ for different (pre-synaptic) connectivity levels, exhibiting a $\sim 1/K$ dependency.

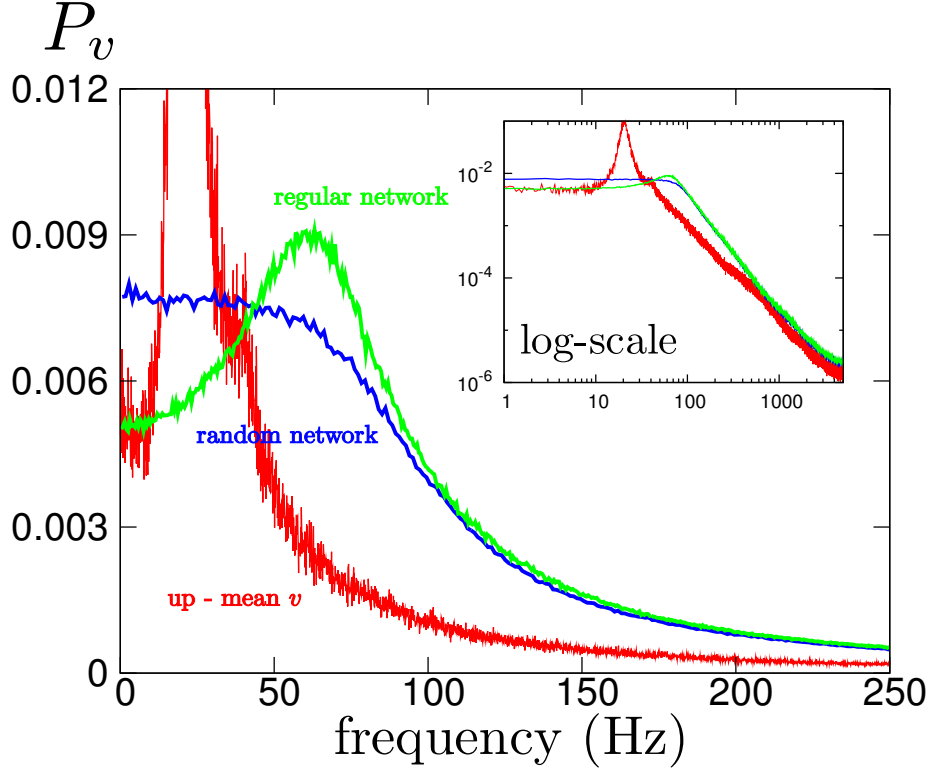

**Figure S5-2.** Averaged power spectra for individual neuron membrane-potential time-series for both a random network with average connectivity $K = 7.5$ (blue) and a regular network with connectivity $K = 7$ (green) in the Up state. A sharp peak (around 60 Hz) is seen for regular networks; instead in random networks the peak is blurred owing to node-to-node heterogeneity. In any case, there is no peak at the characteristic frequency of the global, network-averaged membrane potential, $\approx 20$ Hz (peak of the red curve): individual neurons do not lock to the collective rhythm within Up states. The inset shows a similar plot in logarithmic scale, putting forward the presence of a distinct peak for regular networks together with a $w^{-2}$ tail for all spectra.

## S6 Power-spectrum evaluation for Model B

To firmly establish the correspondence between the phenomenology described for Model B in the main text and stochastic amplification we need to write down a set of effective Langevin equations (analogous to equation 1 in the main text) for the network averaged variables and, from it, compute power-spectra. This turns out to be a non-trivial task.

Our starting point is equation (s7) above. Multiplying it by $V$ and integrating over all possible values of the membrane potential variable

$$
\begin{aligned}
\dot{v}(t) &= \int_{V_r}^{\theta} V \frac{\partial P(V,t)}{\partial t} dV = VD \frac{\partial P(V,t)}{\partial V}\bigg|_{V_r}^{\theta} - D \int_{V_r}^{\theta} \frac{\partial P(V,t)}{\partial V} dV \\
&\quad - V\nu_d(V) P(V,t)\bigg|_{V_r}^{\theta} + \int_{V_r}^{\theta} \nu_d(V) P(V,t) dV \\
&= \theta D \frac{\partial P(\theta,t)}{\partial V} - V_r D \frac{\partial P(V_r,t)}{\partial V} + DP(V_r,t) + V_r \nu_d(V_r) P(V_r,t) + \int_{V_r}^{\theta} \nu_d(V) P(V,t) dV \\
&= -(\theta - V_r)f(t) - \frac{v - V_r}{RC} + V_e f_e + KuV_{\text{in}}f(t) + DP(V_r,t), \quad \text{(s16)}
\end{aligned}
$$

where boundary conditions have been imposed and $\tau_{rp}$ has been, for simplicity, neglected.

The self-consistent method used for solving equation (s7) together with equation (s9) provides $v^*$, $u^*$, $f^*$ and $P(V_r)^*$, computed via the mean, the slope in $\theta$ and the value in $V_r$ of the steady state solution $P(V)$. Results are shown in Table S6. Differences with simulation results ($\langle v \rangle_{\text{up}} = -61.67\,\text{mV}$, $\langle u \rangle_{\text{up}} = 0.2352$; $\langle v \rangle_{\text{down}} = -68.3\,\text{mV}$, $\langle u \rangle_{\text{down}} = 0.997$) stem from the relatively small value $n_r = 6$ used in simulations; as explained above, the Fokker-Planck approach is strictly valid for $n_r \to \infty$.

|  | **Up** ($p_r = 0.5$) | **Down** ($p_r = 0.2$) |
|---|---|---|
| $v^*$ (mV) | -61.16 | -66.54 |
| $u^*$ | 0.2108 | 0.999996 |
| $f^*$ (Hz) | 74.88 | 0.000216 |
| $P(V_r)^*$ (V$^{-1}$) | 40.75 | 150.04 |

**Table S6.** Results obtained from the Up and Down steady state distributions shown in Fig. S2.

In the case of finite networks, $P(V_r, t)$ and $f(t)$ become fluctuating time-dependent variables. Fig. S6-1 illustrates the result of numerical simulations for network of various sizes for the firing rate: $f(t)$ is observed to be strongly correlated with $v(t)$; the larger the average potential the larger the fraction of neurons firing per unit time. The inset of Fig. S6-1, where $f$ is plotted as a function of $v$ and $u$, illustrates the existence of two well-defined branches, one for up-to-down transitions and another for down-to-up, when $f$ is considered a function of $v$.

For the forthcoming analytical calculations, $f$ can be well approximated by a threshold-linear (or "split") function of $v$ plus a noise, and with this we describe its shape in both Up and Down states. The noise amplitude, as shown in Fig. S6-2 decreases with network-size as expected on the central limit theorem basis.

As for the probability density of neurons at the resting state, $P(V_r, t)$, it is consider as a constant of value $P(V_r)^*$ for simplicity.

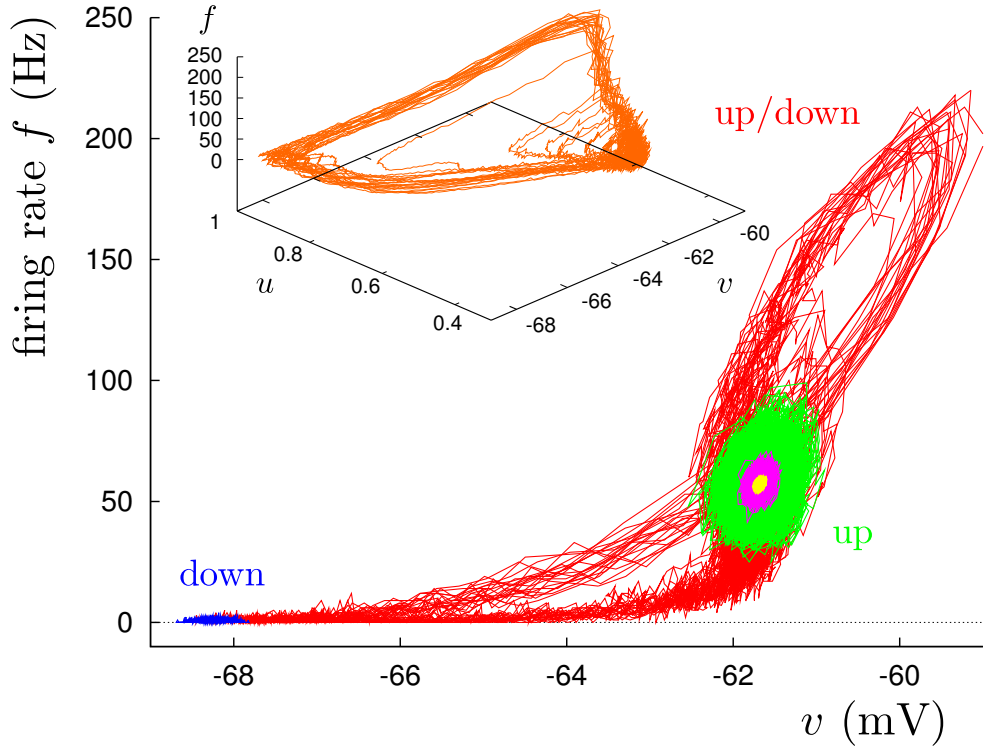

**Figure S6-1.** Main: Firing rate $f$ as a function of the mean membrane potential $v$ for various finite-size networks in the model of Millman *et al.* for (red) the Up-and-Down state $p_r = 0.3$ and $N = 10^3$, (blue) Down-state $p_r = 0.2$ and $N = 10^3$, and (green $N = 10^3$, magenta $10^4$, and yellow $N = 10^5$) Up-state $p_r = 0.5$. By increasing the system size the cloud of points converges to the steady state fixed point. The approximate linear fit is:
$f_{\text{up}}(v) = (12.86 \pm 0.05\,\text{Hz/mV})v + (850 \pm 3)$Hz for the Up-state and $f(v) = 0$ for the Down-state. Even in the case of Up-and-Down states, $f(v)$ can be well approximated by a bi-valuated function with two branches: one for transitions Down-to-Up and other for Up-to-Down. Inset: $f$ as a function of both $v$ and $u$ illustrating the origin of the two branches above.

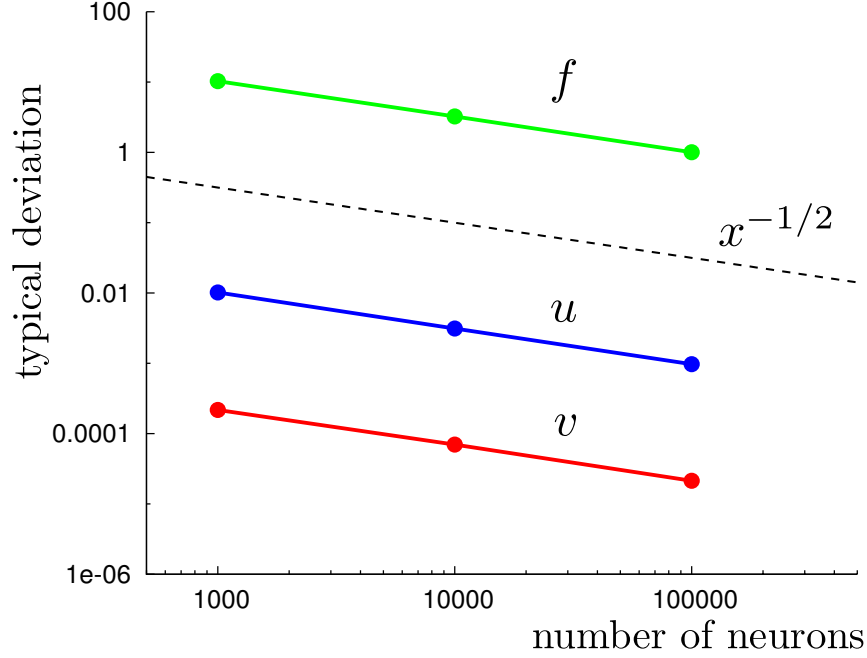

**Figure S6-2.** Typical deviation of fluctuations for different variables as a function of the system size. Simulations -computed for a persistent Up state– show a decay of $1/\sqrt{N}$, as expected on the basis of the central limit theorem.

Having an analytical approximation for $f(v)$ it is now possible to perform a lineal stability analysis. Defining $x = v - v^*$ and $y = u - u^*$ as the linear deviations from the deterministic fixed points, the corresponding Jacobian matrix is specified as follows:

$$\begin{aligned}
a_{vv} &= -\frac{1}{RC} + f' \left\{ n_s u \bar{V}_{in} \left(1 + \frac{1}{2} u \bar{V}_{in} P(V_r)\right) - (\theta - V_r) \right\} \\
&\quad + G \left\{ \bar{V}_e f_e + n_s f u \bar{V}_{in} + 2DP(V_r) \right\} \\
a_{vu} &= n_s \bar{V}_{in} f \left\{ 1 + u \bar{V}_{in} P(V_r) \right\} \\
a_{uv} &= -u p_r f' \\
a_{uu} &= -\frac{1}{\tau_R} - p_r f
\end{aligned} \quad (s17)$$

where $G$ is the derivative of the re-scaling factor of the incoming currents (see S7) which depends

on $f(v)$:

$$G \equiv \frac{\tau_s f' \left\{ e^{-\frac{1}{f\tau_s}} \left(1 + \frac{1}{\tau_s f}\right) - 1 \right\}}{1 - f\tau_s \left\{ 1 - e^{-\frac{1}{f\tau_s}} \right\}} \tag{s18}$$

giving a non-trivial correction.

At the Up-state fixed point this leads to $a_{vv} = -120.12$ Hz, $a_{vu} = 10.4272$ V·Hz, $a_{uv} = -1355.44$ Hz/V, $a_{uu} = -47.4422$ Hz for the coefficients of the stability matrix, and hence a minimum at the denominator of $P(w)$ at $w_0 = 76.1$ rad/s $\Longrightarrow f_0 = 12.11$ Hz. Instead, in the Down-state, the equation for $u$ becomes decoupled from that for $v$, resulting in the absence of a non-trivial peak in the spectrum (complex $\omega_0$), even for a small but non zero firing rate.

## S7 Re-scaling of the incoming currents

When a neuron fires, it is kept silent during the refractory period and it "ignores" all the arriving currents. As the tails of exponential (internal or external) currents can be interrupted by this mechanism, taking the mean value of the exponential function (integrated between 0 and $\infty$) is not a good approximation. A better estimation can be obtained as follows. In the mean-field approach, a neuron fires every $1/f$ seconds; meanwhile, the incoming currents contribute to increment its membrane potential; a schematic representation is shown in Fig. S7.

Therefore, one can consider that the neuron integrates on average from the spiking-time $t = 0$ to some effective final time $t = 1/f$. The contribution of an incoming spike arriving at $t = t_s$ can be computed as

$$\overline{V}_{\text{e/in}}(t_s) = \int_{t_s}^{f^{-1}} w_{\text{e/in}} e^{-\frac{t-t_s}{\tau_s}} dt = V_{\text{e/in}} \left[ 1 - e^{-\frac{f^{-1}-t_s}{\tau_s}} \right], \tag{s19}$$

where $V_{\text{e/in}} = w_{\text{e/in}} \tau_s / C$. As many spikes arrive during the interval $[0, 1/f]$, supposing an uniform

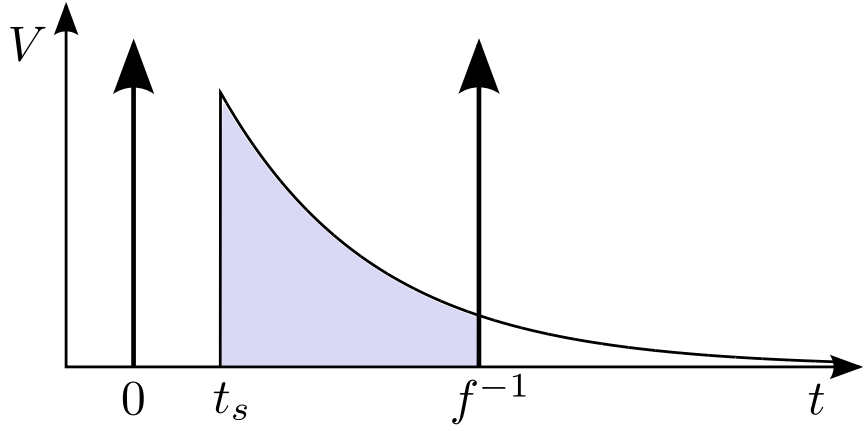

**Figure S7.** Representation of an incoming, exponential current integrating until the post-synaptic neuron fires, entering into the refractory period. Losses depend on the spiking time of the pre-synaptic neuron and the firing rate of the post-synaptic one.

distribution for the incoming times, the mean value of this $t_s$-function is

$$\bar{V}_{e/in} \equiv \frac{\int_0^{f^{-1}} \bar{V}_{e/in}(t_s)dt_s}{f^{-1}} = V_{e/in}\left[1 - f\tau_s\left(1 - e^{-\frac{1}{f\tau_s}}\right)\right]. \quad (s20)$$

Observe that an extra factor multiplying the mean value of the exponential has appeared with respect to the naive estimation in [s6]. For typical values of $f \approx 100$ Hz and $\tau_s = 5$ ms, the mean contribution of incoming spikes is rescaled by a more than a 50%.